\PassOptionsToPackage{table,xcdraw}{xcolor}

 \documentclass[manuscript,screen]{acmart}

\usepackage{caption}
\usepackage{subcaption}
\usepackage{enumerate}
\usepackage{adjustbox}
\usepackage{tabulary}
\usepackage{array}
\usepackage[table,xcdraw]{xcolor}
\usepackage{todonotes}

\begin{document}

\title{ODSearch: Fast and Resource Efficient On-device Natural Language Search for Fitness Trackers' Data}

\author{Reza Rawassizadeh}
\email{rezar@bu.edu}
\affiliation{%
  \institution{Department of Computer Science, Metropolitan college, Boston University}
  \country{United States}
}
\author{Yi Rong}
\email{yrong@bu.edu}
\affiliation{%
  \institution{Department of Computer Science, Metropolitan college, Boston University}
\country{United States}
}

\authornote{Both authors contributed equally to the paper.}

\begin{abstract}
Mobile and wearable technologies have promised significant changes to the healthcare industry. Although cutting-edge communication and cloud-based technologies have allowed for these upgrades, their implementation and popularization in low-income countries have been challenging. We propose \emph{ODSearch}, an On-device Search framework equipped with a natural language interface for mobile and wearable devices. To implement search, \emph{ODSearch} employs compression and Bloom filter, it provides near real-time search query responses without network dependency. In particular, the Bloom filter reduces the temporal scope of the search  and compression reduces the size of the data to be searched. Our experiments were conducted on a mobile phone and smartwatch. We compared \emph{ODSearch} with current state-of-the-art search mechanisms, and it outperformed them on average by \emph{53 times in execution time, 26 times in energy usage, and \emph{2.3\% in memory utilization} }.
\end{abstract} 


\maketitle

\section{Introduction}
Technological advancements and widespread adoption of mobile and wearable technologies has transformed healthcare services and improved the quality of delivery by offering convenient low-cost apparatuses that provide inexpensive health monitoring and prognosis  \cite{guo2015verifiable, dananjayan20215g,hathaliya2020securing,zhang2021passive,liao2018just}. Consequently, new scientific disciplines have arisen, such as mobile health (mHealth) or employing wearables for symptom diagnosis, and their popularity has increased from the Covid-19 outbreak \cite{gan202111,adans2020can,quer2021wearable}. As a result, the demand for mHealth integration into the current healthcare system is also increasing \cite{osei2021mobile}. One primary application of mHealth is operating the devices as a data collection and repository for Personal Health Records (PHRs) \cite{rawassizadeh2013ubiqlog,dobbins2017detecting,jiang2019memento} that are either manually entered by users \cite{bouayad2017patient}, or automatically collected by device sensors, such as physical activities \cite{tang2018defining,jourdan2020privacy} and heart rate \cite{mohamed2017heartsense}. 

mHealth services provide affordable access to medical facilities, especially in low-income countries with limited resources \cite{aranda2014systematic}, which promises to improve the quality of healthcare in those regions. However, despite continuous efforts, mHealth has not yet been extensively adapted \cite{colaci2016mhealth,aref2019factors}.

Implementation of mHealth applications in low-income countries is associated with three significant challenges: (i) network availability, (ii) privacy and security, and (iii) literacy level and IT skills. 

\textit{Network availability:} a large number of the mHealth applications are designed based on expensive communication technologies, such as 5G networks \cite{dananjayan20215g}, Blockchain \cite{taralunga2021blockchain,liu2021blockchain}, and Cloud Services \cite{guo2015verifiable}, and require integration of complex resource-intensive machine learning algorithms into their applications \cite{keshavarz2020sefr}. The “network” is a barrier to large-scale deployment of mHealth applications in low-income countries, since deployment of mHealth is highly dependent on Internet and local human resources for maintenance. Coverage, accessibility of networks, and Internet availability are not guaranteed in all areas of low-income countries \cite{rawassizadeh2018nocloud,aranda2014systematic}. 

\textit{Privacy and Security:} The  major challenge related to cloud services is “security and privacy,”  such as unauthorized data alteration, unauthorized data access, and unauthorized sharing \cite{iwaya2020security,alam2020mhealth}. This is a significant issue in low-income countries that often suffer from authoritarian governments and lack proper regulation to prevent access to users' digital information \cite{gobel2013information}.

\textit{Literacy and IT skills:} Low IT skills and literacy is another hurdle required to adapt mHealth technologies in low-income countries \cite{koopman2020conversational,paglialonga2019mhealth}. Studies reveal that literacy is directly correlated with the Gross Domestic Product (GDP) of a country \cite{mehmood2014health}, and low-income countries suffer from literacy and low IT skills.

Users with the described features could benefit from mHealth applications, by \emph{keeping all the computation and data on-device and via Natural Language Interfaces (NLI), facilitating communication with applications}. By using NLI, even users with low literacy and low IT skills can benefit from mobile applications as well \cite{koopman2020conversational}. NLI uses natural language to interact with an application and can solve the literacy problem that Graphical User Interfaces (GUI) present when interacting with mHealth applications \cite{liu2017toward,rawassizadeh2017natural}. Traditional Information Retrieval (IR) systems are based on distributed servers or resource-intensive processes \cite{hersh2020information}. To enable reliable user interaction with an mHealth application, we introduce an alternative on-device approach.

Here we propose \textit{ODSearch} \footnote{To allow full reproducability, our codes are available at https://drive.google.com/drive/folders/1b6Rj1Vgdinr0WnABNk2NqXFGYoluab5a?usp=sharing}, 
an \underline{O}n-\underline{D}evice \underline{Search} framework for mobile and wearable fitness trackers with NLI. \textit{ODSearch} enables mHealth applications to search for users' Personal Heath Records (PHRs) using natural language, completely independent from the network itself. Here, by PHR we refer to data from fitness-trackers, that can measure users' heart rate and physical activities.

Several promising approaches have been proposed including an on-device natural language query interface \cite{rawassizadeh2017natural}, on-device classification \cite{rawassizadeh2016scalable,keshavarz2020sefr}, and on-device clustering \cite{rawassizadeh2019indexing}. However, the IR component, in particular the search function, is still absent; our work aims to bridge this gap. \textit{ODSearch} consists of a pipeline of four modules: \emph {Query Translator} bridges the language gap between the natural language query and the machine query, \emph{Bloom Filter}, which queries the membership with efficient search space, \emph{Compression} (i.e., Huffman Coding) which compresses and decompresses the stored dataset and thus facilitates the search. Data compression is the process of converting data from one format to another format that is smaller and therefore takes up less space, which can also shorten the execution time of the search. The last component is \emph{Answer Translator} which presents the results in the natural language, computed from the machine query results.

This research provides the following novel contributions: 
\vspace{-1mm}
\begin{itemize}
    \item We employ two algorithms (Bloom-filter and Huffman coding) to execute on-device data query and search, in \textit{real-time or close to real-time.} Our approach outperforms all state-of-the-art methods in response time and battery utilization, and outperforms most for memory utilization.
    \item We propose a \textit{fast and resource-efficient} search framework that can operate on small devices, even smartwatches. To our knowledge, information retrieval mechanisms on wearables have not been extensively studied. Our experiment shows that state-of-the-art mobile data storage and information retrieval approaches are not suitable for use in small wearables (e.g. smartwatches).
    \item We \emph{quantify the scalability limitations} of current, state-of-the-art methods for searching data on battery-powered devices, including smartwatches and smartphones.  
    \item We \emph{compare compression rate and average encoding/decoding time} for lossless compression algorithms for use on battery powered devices.
\end{itemize}

\section{Related Work}
This section reviews studies that  mitigate IR challenges for mobile and wearable devices. We have organized them into three categories: (i) indexing and information retrieval on resource-constrained devices, (ii) on-device databases and search  frameworks, and (iii) text compression algorithms. 

\subsection{Efficient Information Retrieval and Indexing}

Mobile IR and indexing focus on finding appropriate ways to help users analyze collected contextual data \cite{tsai2010introduction}. Although there are advances to on-device machine learning \cite{dhar2021survey}, extensive exploration of mobile IR is still lacking. 

\textit{Information Retrieval:} There are some promising resource-efficient approaches for IR systems such as, Gupta et al. \cite{gupta2015new} who demonstrated a term-weighting schema-based ranking function, where ranking function was based on fuzzy logic to improve the accuracy of retrieving relevant documents. Subhashini and Kumar et al. \cite{subhashini2011framework} concentrated on improving search accuracy through optimizing natural language processing techniques, by considering only nouns and verbs. Rawassizadeh et al. \cite{rawassizadeh2017natural} proposed an on-device natural language query interface that can parse closed domain queries, but it does not include information retrieval. These studies did not investigate the issue of searching large amounts of data on small devices.

\textit{Indexing:} Another common approach to facilitate access to large-scale data is indexing. Białecki et al. \cite{bialecki2012apache} proposed Apache Lucene \footnote{https://lucene.apache.org}, a full-text search engine, that uses an inverted index as a kernel and is widely used as a practical platform in the industry. Yang et al. \cite{yang2017anserini} proposed 'Anserini', an extension toolkit on top of Lucene, that adds a scalable inverted index, streamlined IR, and an architecture enabling multi-stage ranking. Tan et al. \cite{tan2010microsearch} designed and implemented an upgraded inverted index combined with a hash function to simplify the inverted index that makes the corresponding query more accurate and memory efficient. Rawassizadeh et al. \cite{rawassizadeh2019indexing} used spatio-temporal clustering as index construction to facilitate search and reduce the search space by leveraging spatio-temporal indices.

We chose to compare our approach with Lucene \cite{bialecki2012apache}, because it is widely used as an inverted index on different platforms. We also benefited from a descriptive solution to design our search framework by employing the temporal filtering mechanism used by Rawassizadeh et al. \cite{rawassizadeh2019indexing} to reduce the search space. 

\subsection{On-device Databases and Search Engines}
Tan et al. \cite{tan2010microsearch} built one of the earliest resource-efficient search engines that could work on embedded devices with a top-k query algorithm, by using a buffer cache and an inverted index. Their approach is limited to keyword search and not range queries. Lyu et al. \cite{lyu2017empirical} proposed an empirical study that local databases often employ as the on-device IR system to provide users data storage and retrieval. They also analyzed and identified that the most frequently used databases in Android are SQLite, Oracle, and Realm. However, continuously using the local databases can lead to excessive power consumption or security problems \cite{lyu2017empirical}. Moreover, there are other local databases that can operate on mobile phones and smartwatches with exceptional performance, such as the H2  \footnote{https://www.h2database.com/html/main.html}, LevelDB \cite{dent2013getting}, and ObjectBox \cite{walachowski2020comparative}.

We chose to compare our approach to  three popular databases that run locally on Android devices: SQLite\footnote{https://developer.android.com/reference/android/database/sqlite/package-summary}, Realm\footnote{https://docs.mongodb.com/realm/sdk/android/}, and H2 databases.  From a technical perspective, H2 and SQLite both use a combination of B-tree index and brute force method for searching the data, while Realm only uses B+Tree indices. Therefore, from a lower-level perspective, we are comparing our approach with B-Tree and B+Tree indices. To our knowledge, there is no empirical analysis for databases or search engines that can search data collected by the smartwatch. We compared the execution time, memory utilization, and energy usage in the search tasks on mobile phones and smartwatches of \textit{ODSearch} against the aforementioned listed databases.

\subsection{Text Compression Algorithms}
There are two types of compression algorithms: lossy and lossless. No information is lost after conversion in lossless algorithms. For example, compressing an image from TIFF format to PNG is lossless compression. On the other hand, lossy compression algorithms have a higher compression rate but also lose some information. For example, compressing an image from TIFF format to JPEG is lossy compression. The majority of text compression algorithm are lossless compression algorithms and they can be classified as either statistical-based approaches or dictionary-based approaches \cite{shanmugasundaram2011comparative,welch1984technique}. 

Statistical-based approaches take advantage of each character's frequency and some popular examples are Run-length encoding \cite{golomb1966run}, Shannon-Fano encoding \cite{shannon1948mathematical,fano1949transmission}, Arithmetic encoding \cite{pasco1976source}, and Huffman encoding \cite{huffman1952method}. Run-length encoding can effectively compress consecutive repeated characters in a text. Shannon-Fano encoding can construct prefix code to compress symbols based on their probabilities, but it is challenging to achieve optimal compression efficiency \cite{shanmugasundaram2011comparative}. Arithmetic encoding can approximate the optimal compression ratio, but encoding and decoding are very time-consuming \cite{rahman2020burrows}. Huffman encoding can approach the optimal compression rate if the frequencies of characters are very large \cite{rahman2020burrows}. Huffman encoding is also used in recent federated learning architectures \cite{malekijoo2021fedzip} for reducing communication with the server while transferring neural network weights back and forth. 

Conversely, dictionary-based approaches store all recurring patterns, including single characters and strings of different lengths, while keeping the mapping relation between patterns and their codes in a dictionary. This makes dictionary-based approaches relatively efficient, but their search process is computationally expensive \cite{rahman2020burrows}, possibly resulting in excessive encoding and decoding time. A popular example, Lempel–Ziv–Welch (LZW) compression, is used in Linux based operating systems \cite{duricek2015hybrid}.

We compared listed lossless compression later and report our rationale for our decision.

\section{Methods}

\begin{figure*}
        \centering
        \begin{subfigure}{\textwidth}
            \centering
            \includegraphics[width=\textwidth]{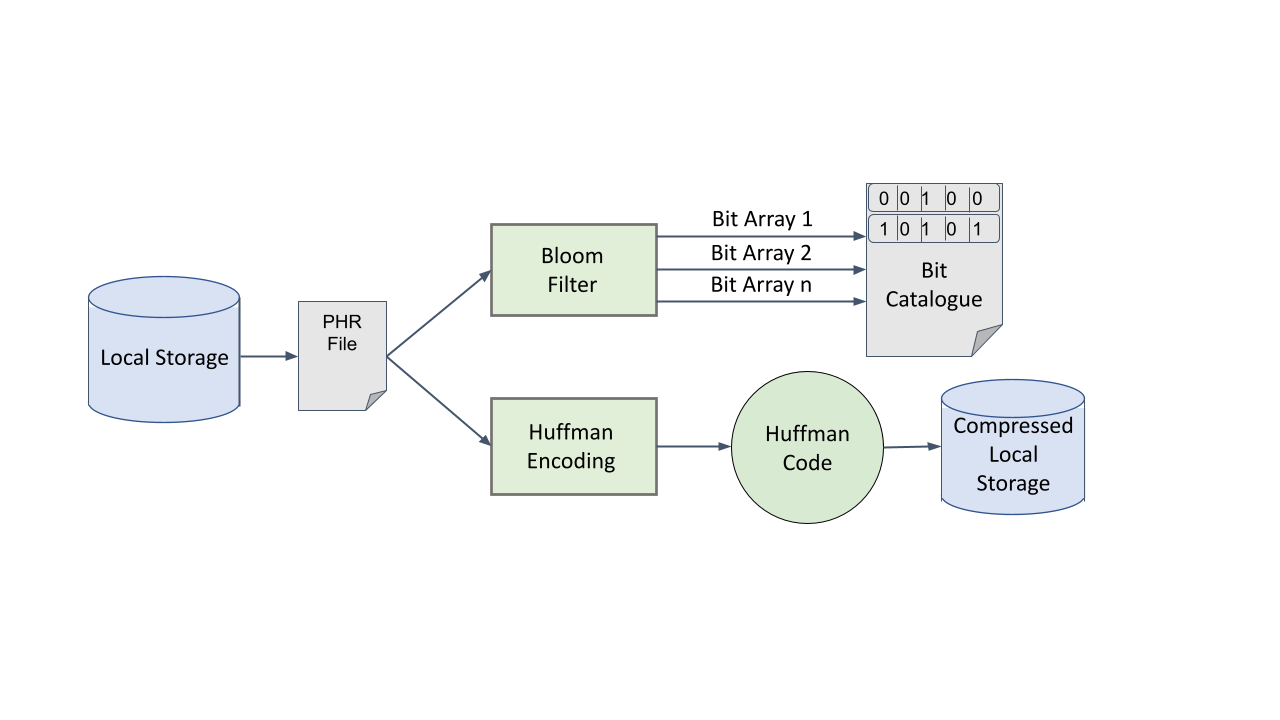}
            \vspace{-25mm}
            \caption{\small{Phase 1: Preprocessing}}
            \label{fig:1-1}
        \end{subfigure}
        \begin{subfigure}{\textwidth}   
            \centering 
            \includegraphics[width=\textwidth]{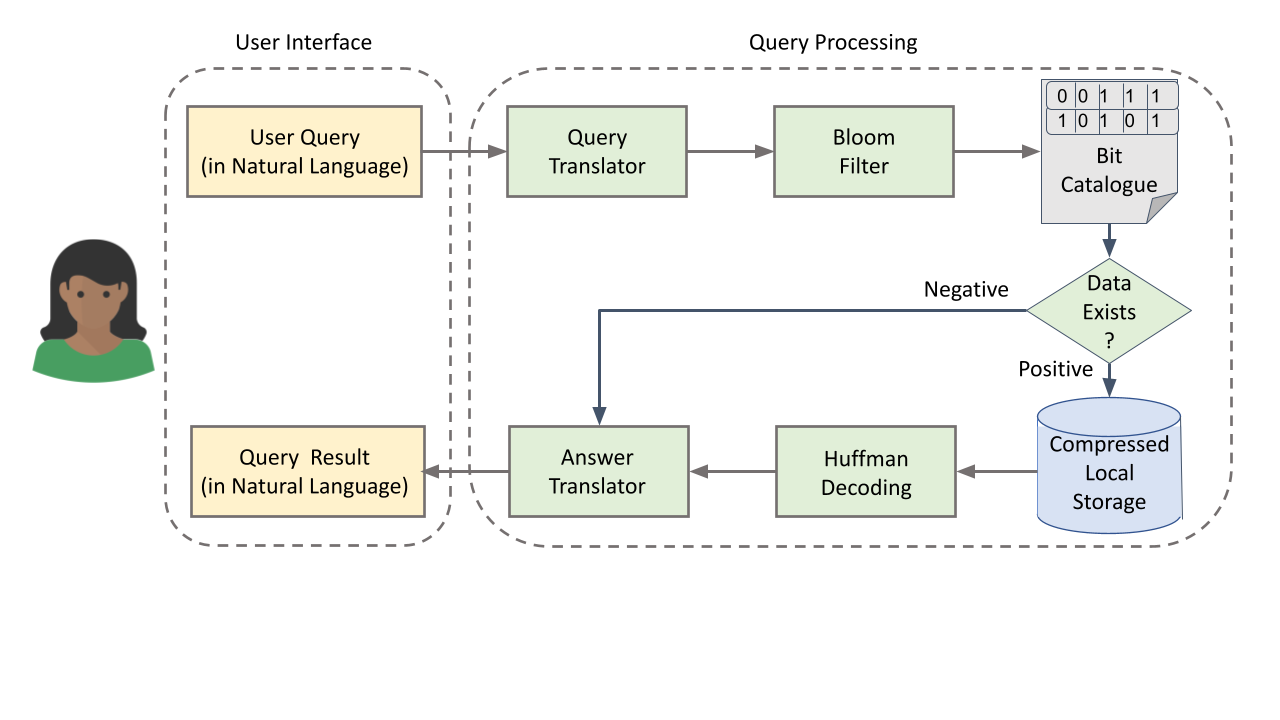}
            \vspace{-2cm}
            \caption{\small{Phase 2: Query execution}}    
            \label{fig:1-2}
        \end{subfigure}
        \hfill \hfill
        \caption{\small{Two phases of \emph{ODSearch} process flow}} 
        \label{fig:1}
\end{figure*}

IR algorithms refer to algorithms used for indexing and retrieving structured or unstructured information from text, video, images, and audio. Hersh et al. \cite{hersh2020information} explained that four modules are required for building an IR system for health applications: content, metadata, search engine, and queries. In line with those principles, \textit{ODSearch} includes four modules: two focused on searching the content and two focused on connecting natural language queries with the search engines. Those modules are (i) Query Translator, (ii) Bloom Filter, (iii) Compression, and (iv) Answer Translator. Among these modules, \textit{ODSearch} builds a pipeline of query, search, and information retrieval, and operates in two phases shown in Figure \ref{fig:1} (a) and (b). The first phase, “Preprocessing”, focuses on constructing indices for the underlying datasets. It operates periodically, such as once per day, and processes the new data that was added to the system. This phase transfers the raw data from local storage to the Bloom filter and compresses it with Huffman encoding. Bloom filter constructs bit arrays for each day's data entries, which we call the “Bit Catalogue”. Huffman encoding performs the compression on the original data, and stores it in “Compressed Local Storage” (CLS).

The second phase is "Query Execution". In this phase, the framework uses the “Query Translator” to obtain keywords from the natural language query entered by the user. Next, query keywords are searched in Bit Catalogue. If a positive result is returned (i.e., keywords may exist in the content), then the query keywords will be subsequently searched in the CLS. If the keywords do not exist and the result of the Bit Catalogue check is negative, subsequent searching in the CLS and “Huffman Decoding” will be skipped. Finally, the “Answer Translator” calculates all search results from the previous step and presents the final query result in natural language. The term “calculate” here, refers to summation, minimum, maximum, and average. Figure \ref{fig:1} shows the architecture of the \emph{ODSearch} framework. We have separated the Preprocessing phase from Query Execution for clarity. Each component of Figure \ref{fig:1} will be explained later in more detail.

With this architecture,  we were able to get a near real-time response even with very large amounts of data on a resource-constrained device such as a smartwatch.

\subsection{Datasets}

To conduct our experiments, we used two real-world datasets, one from smartphones and one from smartwatches. We used the “Ubiqlog” dataset \cite{rawassizadeh2013ubiqlog}, generated from a lifelogging tool on mobile phones, and the “Insight for Wear“ dataset \cite{rawassizadeh2015energy}, generated from a continuous sensing tool on a smartwatch. Both UbiqLog\footnote{https://archive.ics.uci.edu/ml/datasets/UbiqLog+(smartphone+\symbol{92}lifelogging)} and Insight for Wear\footnote{https://play.google.com/store/apps/details?id=com.insight.insight} datasets are publicly available for research purposes. 

Mobile and wearable devices can provide robust and affordable means to access data regarding a users’ heart rate, step count, and activity type \cite{georgiou2018can,gholamiangonabadi2020deep,huang2016validity,johnston2021recommendations}. Moreover, these are the de-facto standard information objects generated from mobile and wearable device sensors, and they are among the most frequently asked values when users query their PHRs \cite{rawassizadeh2017natural}. We use heart rate, step count, and activity type from the  "Insight for Wear" dataset and the step count and activity type data from the "UbiqLog" dataset. Missing data due to a variety of sensors within different devices is inherent in these datasets \cite{rawassizadeh2019ghost}. Therefore, we chose the user from each dataset (two users total) with the most available data and consequently the fewest missing data points. Choosing the largest dataset available simulates the worst case of a real world search. 

Original PHR data collected from mobile or wearable sensors are all associated with time. A PHR record, \textit{r}, can be represented within a 3-tuple arrangement: \(r = <S, T, D>\). \textit{S} denotes the sensor name, \textit{T} is the record's timestamp, and \textit{D} is the sensor data. We define the combination of \textit{S, T, D} as the metadata of the content. Among the three categories of the dataset, heart-rate data is recorded as beats per minute (bpm), step count refers to the total number of steps taken in a day. The activity types, including “still”, “tilting”, “onfoot”, “invehicle”, and “unkonwn”, were extracted by the Google Fit library \footnote{https://developers.google.com/fit}. The following are examples of each sensor reading in JSON format:

\texttt{
\hspace{-6mm}\{"sensor\_name":"HeartRate","timestamp":"Mon Apr 17 15:00:35 EDT 2017","sensor\_data":\{"bpm": 60\}\}\\
}
\texttt{
\{"sensor\_name":"ActivFit","timestamp":"Thu Apr 06 23:50:24 EDT 2017","time\_stamp":"Thu Apr 6 23:50:24 EDT 2017","sensor\_data":\{"step\_counts":755,"step\_delta":13\}\}\\
}
\texttt{
\{"Activity": \{"start":"4-2-2017 19:41:00","end": "4-2-2017 19:41:00","type":"onfoot","condfidence":\\"80"\}\}. \\
}
To demonstrate the scalability of our approach, we created synthetic datasets based on the two real users we selected. The synthetic data is simply a repetition of the original data segment, ignoring the data distribution. The size of the synthetic segments varied between 30\;Kb and 48\;Mb. According to the number of records, the largest synthetic dataset simulates approximately over three years worth of data. Moreover, the upper limit of 48\;Mb was based on the processing capacity of both devices. Table \ref{tab:1} provides a summary of the number of instances that occured in the real-world and synthetic datasets.

\begin{table}
    \begin{subtable}{\textwidth}
        \centering
\begin{tabular}{|l|r|r|}
\hline
Dataset   Size & \multicolumn{1}{c|}{\begin{tabular}[c]{@{}c@{}}Experimental Sample of \\ One User\end{tabular}} & \multicolumn{1}{c|}{\begin{tabular}[c]{@{}c@{}}Original Real-world\\ Datasets\end{tabular}} \\ \hline
Heart   Rate   & 491  & 375,832   \\ \hline
Step Number    & 1,620  & 1,571,081    \\ \hline
Activity Type  & 8,154   & 167,951  \\ \hline
Total          & 10,265  & 2,114,864       \\ \hline
\end{tabular}
       \caption{Real-world datasets}
    \end{subtable}
    \hfill
    \begin{subtable}{\textwidth}
        \centering

\begin{tabular}{|l|r|r|r|r|r|r|r|r|r|r|r}
\hline
Dataset   Size & 30kB & 3MB  & 6MB  & 12MB & 18MB & 24MB & 30MB & 36MB & 42MB & 48MB\\ \hline

Heart Rate     & 108 & 10,314  & 20,622 & 41,238 & 61,860 & 82,476 & 103,098 & 123,714 & 144,330 & 164,952\\ \hline
Step Number    & 66 & 6,066 & 12,126 & 24,246 & 36,366  & 48,486 & 60,612  & 72,732& 84,852  & 96,972 \\ \hline
Activity Type  & 96 & 9,234 & 18,462 & 36,924 & 55,386 & 73,848 & 92,310 & 110,772    & 129,234 & 147,696 \\ \hline
Total   & 270 & 25,614 & 51,210 & 102,408 & 153,612 & 204,810 & 256,020 & 307,218 & 358,416 & 409,620 \\  \hline

\end{tabular}

        \caption{Synthetic datasets}
     \end{subtable}
     \caption{Quantitative summary (number of instances) of datasets}
     \label{tab:1}
\end{table}

\subsection{Bloom Filter}
The Bloom filter \cite{debnath2011bloomflash, luo2018optimizing} is a space-efficient probabilistic data structure used to search for member existence within a dataset. A Bloom filter consists of an array of \textit{m} bits, which can represent \textit{n} elements in a set $S = \{x_1, x_2, . . . , x_n\}$. Initially, all bits are set to zero. In the process of inserting elements into a database that has a Bloom filter, each element \(x \in S\) is hashed with \textit{k} hash functions $(h_{i}(x), 1 \leq i \leq k)$, and mapped to random numbers uniformly distributed among the indices of the bit array. In the meantime, the mapped bit $h_{i}(x)$ is set to one. When an element is queried, it may exist in the \textit{S}, only if all the bits $h_{i}(x)$ are set (bits are equal to one) and it does not exist if any of $h_{i}(x)$ bit does not set (bits are equal to zero). Bloom filters are constructed from original datasets, and in the preprocessing phase, their results are saved in the “Bit Catalogue”.

\begin{figure*}[htb]
\begin{center}
\includegraphics[scale=0.3]{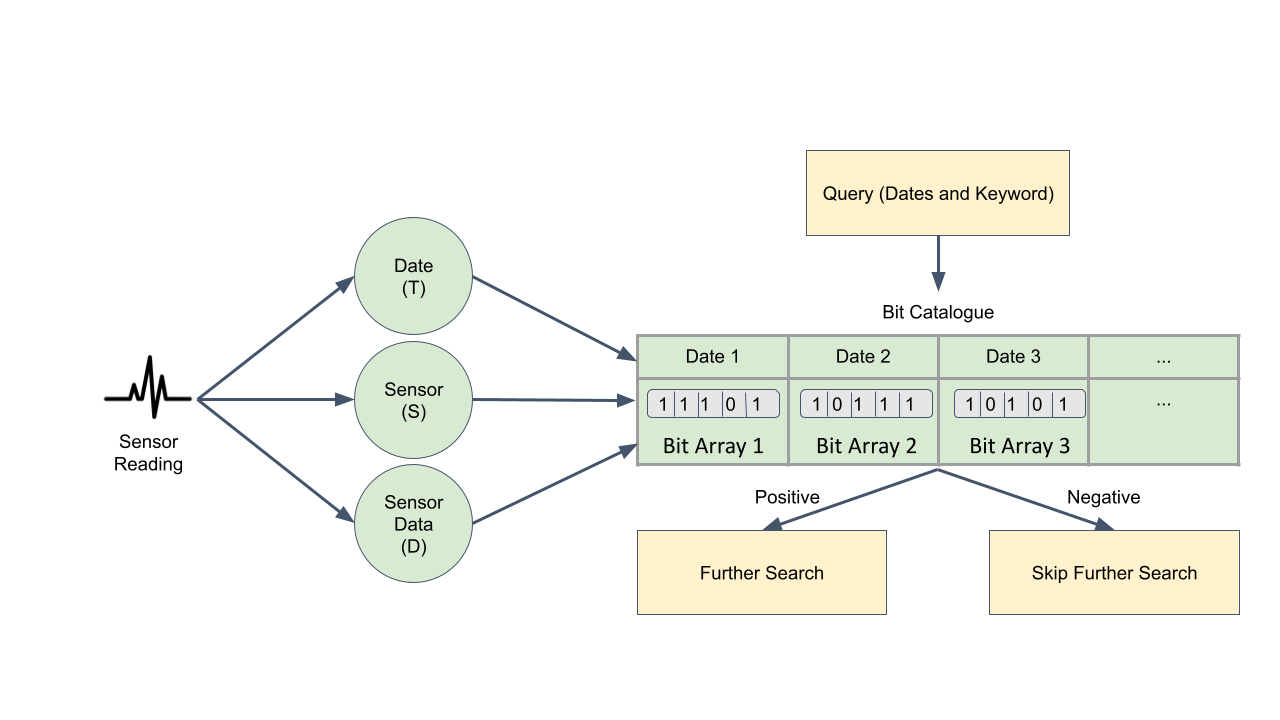}
\caption{ \small Bloom filter module workflow}
\label{fig:2}
\end{center}
\end{figure*}

We implemented Bloom filters through Guava \cite{bejeck2013getting} Java library. To support Bloom filter query based on time, we insert records from different dates to different Bloom filter instances. It means, we created an independent bit array for each day's data. Bit arrays were collected and stored in the "Bit Catalogue". Our "Bit Catalogue" a key-value pair data structure whose keys are unique, used date as a key and a bit array as value. 

The process of implementing the Bloom filter module is shown in Figure \ref{fig:2}. First, we extracted the three metadata attributes \textit{S, T, D} from the original data in local storage, then we assigned the date of \textit{T} as the key of the Bit Catalogue, and inserted \textit{S} and \textit{D} into the bit array, which is the corresponding value for the key. This approach makes each pair (date and bit array) unique and independent in the Bit Catalogue. When a user queries data of a certain period, it can retrieve all dates in the given period and use them to look for the corresponding bit array in the Bit Catalogue. If the queried bit arrays return any positive result, the algorithm proceeds with a query to the next step, which fetches all records containing the keyword. If none of the queried Bloom filters returns a positive result, the searched keyword never appears in the given period. Then, no further search will be done. In other words, the Bloom filter module operates as an 'initial filter' before searching the content of data files. This approach enhances the efficiency of the search and enriches the functionalities of the search framework. 

A Bloom filter is prone to false positives because it could claim an element belongs to a set even when it is absent. However, within our framework, getting false-positive results just means the queried element will be further searched in the next step. The evaluation section demonstrates the significant impact of using the Bloom filter.

\subsection{Huffman Encoding and Decoding}

Information storage is not typically an issue with cloud systems, but every query involves delays, both in the cloud and on-device. \cite{rawassizadeh2018nocloud, dhar2019device, zhou2021device}. To reduce the delay, we analyzed several lossless text compression algorithms including, Run-length encoding, Shannon-Fano encoding, Arithmetic encoding, Huffman encoding, and LZW compression \cite{shanmugasundaram2011comparative}. Furthermore, based on the results of our analysis, we chose to use Huffman encoding to compress data and facilitate data retrieval from the encoded data.

Huffman encoding replaces characters that occur more frequently with binary codes that require fewer bits, and replaces less frequent characters with binary codes that require more bits. The process of Huffman encoding starts from counting the frequency of each character, and then a Huffman tree (a type of binary tree) is constructed based on those frequencies. Therefore, a binary representation for each character can be inferred from the Huffman tree. In simple words, Huffman encoding, and the later decoding, are character-based, and suitable for compressing textual content.

\begin{figure*}[htb]
\begin{center}
\includegraphics[scale=0.3]{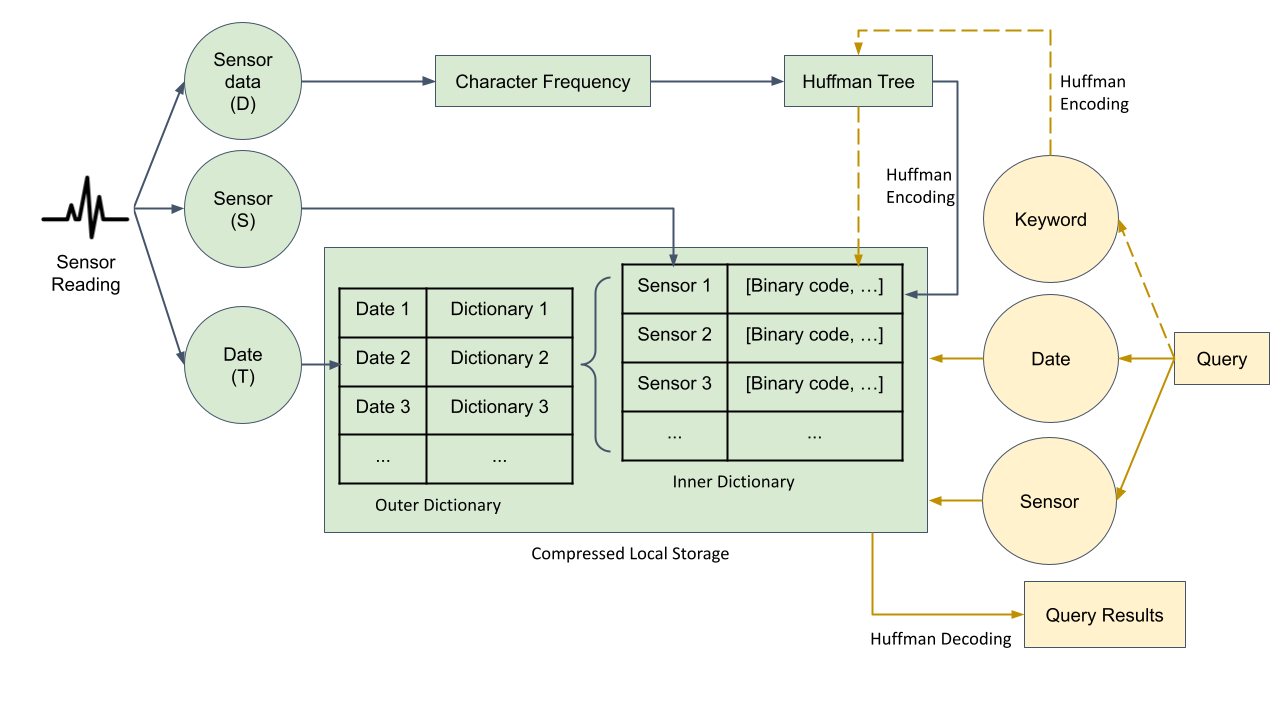}
\caption{ \small Huffman coding module workflow. Green shapes indicate Huffman coding parts, yellow shapes indicate Query parts.}
\label{fig:3}
\end{center}
\end{figure*}

The implementation of Huffman coding in this work is customized from the existing work in \textit{Algorithms} \cite{sedgewick2011algorithms}. The process flow of Huffman coding is shown in Figure \ref{fig:3}. This module begins with counting the frequency of each character in the sensor data, and then it generates a Huffman tree. Since almost all mobile and wearable health related queries are time-dependent \cite{rawassizadeh2017natural}, therefore we compressed the data by time to enable temporal query, and we developed a CLS that included a "Nested Dictionary". Here a Nested Dictionary refers to key-value storage, and it uses a dictionary as a value to allocate data by their dates and sensor types. In other words, the CLS is a two-layer dictionary with a nested structure. The outer dictionary uses the \textit{T} as a key and an inner dictionary uses \textit{T} as a value. It means, we create a new dictionary for each \textit{T}. The internal dictionary assigns \textit{S}, sensor name, as a key and a collection of the sensor data’s binary codes as value. All of the sensor data’s binary codes will be collected in an array and stored in the value of the inner dictionary. For instance, if we refer to an activity type PHR on 08/01/2021 showing “onfoot” as the sensor data, it will be encoded as 001011011100001010. And if we define the content between curly braces as a dictionary (i.e.,  {key: value}) and the content between square brackets as an array, the sensor reading will be stored in the Nested Dictionary as \{“08-01-2021”: \{“activity”: [001011011100001010]\}\}. Inside the Nested Dictionary, “08-01-2021” is the key of the outer dictionary and “activity” is the key of the inner dictionary, and [001011011100001010] is the value of the inner dictionary. Using this compression level, we can process keyword-based queries for sensor data, sensor type, and date. We encoded the query keyword (keyword is extracted from user query) from the constructed Huffman tree and then searched for them in the CLS. After fetching all the compressed results matching the three criteria (date, sensor type, and keyword for sensor data), we used “Huffman Decoding” to translate the query results into human-understandable text. However, if the query does not search for a specific keyword, all records that match the date and sensor type will be retrieved.

We report the difference between using and not using this type of Huffman coding in the "Experimental Evaluation" section.

In Figure \ref{fig:3}, the dotted lines show that the execution of Huffman encoding depends on keyword-based queries and the presence of the keyword. However, since results will always be decoded from the CLS, we do not use a dotted line for Huffman decoding. 
\subsection{Query Translator and Answer Translator}

To enable the NLI, we designed a translation component (Query Translator) between users’ questions (natural-language query) and machine understandable language. “Query Translator” extracts keywords from users’ queries and organizes the keywords for the underlying search. The “Answer Translator” receives the query results, and decorates the results (either numerical or textual) in human-readable language.

Rawassizadeh et al. \cite{rawassizadeh2017natural} proposed that user’s natural language query can be processed with four dictionaries of keywords including: (a) question words, such as “how many'' or “what”; (b) temporal notion, such as “today” or “this week”; (c) sensor name, such as “heart rate” or “activity”; and (d) aggregation words, such as “total” or “average”. However, there is no implementation proposal that executes the query on the underlying dataset. Therefore, we implemented “Query Translator” to bridge this gap, and it implements query classifications proposed by previous works \cite{li2014nalir} to construct the machine query.

\begin{figure*}
\begin{center}
\includegraphics[scale=0.33]{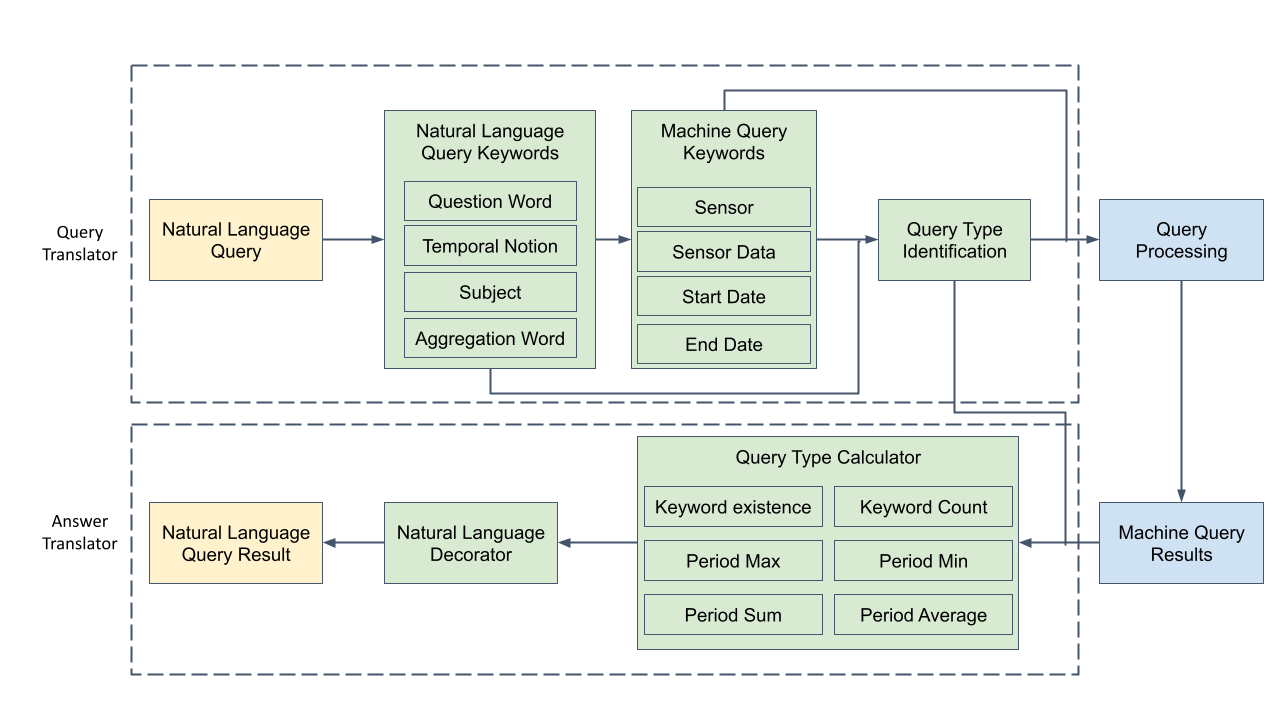}
\caption{ \small Workflow of Query Translator and Answer Translator (yellow shapes indicate user interface, green shapes indicate Query Translator and Answer Translator, blue shapes indicate Bloom filter and Huffman coding)}
\label{fig:4}
\end{center}
\end{figure*}

Figure \ref{fig:4} presents the workflow of the “Query Translator” and “Answer Translator” modules. Query Translator begins by identifying four query keywords: a question word, a subject word, a temporal notion, and an aggregation word. This classification and possible types of questions that users could ask about their PHR, have been proposed in previous works \cite{rawassizadeh2017natural}. 

We identified query types by conducting a manual qualitative analysis of the question dataset proposed in \cite{rawassizadeh2017natural}. Two researchers performed theme analysis and their Fleiss-Kappa score was 82\%, indicating a substantial agreement. The descriptions of query types are shown in Table \ref{tab:2}, and Table \ref{tab:3} presents examples of parsing five different query types. We take the query type to refine essential information of the natural language query and use it to control the “Answer Translator”. For example, one of the query types is “keyword existence”, which means the query only asks for the membership of a keyword; hence only Bit Catalogue will be queried, and subsequent search on the CLS will be skipped. After the keywords and the query type are identified, the "Query Processing" component, which consists of the Bloom Filter and Huffman Coding, will search the preprocessed data and return records matching the query keywords. 
 
Before presenting returned records to users, the “Answer Translator” decorates them and converts them to natural language text. In particular, the “Answer Translator”  performs natural language decoration on the query result, which is coming from the search component. The computed result can be either Boolean or numeric results. In both cases, they are decorated in a human-readable expression, and the final query result is presented on the user interface. 

For example, if a user query is \textit{“How many times did I run this month?”}, the query type will be “keyword count” and the query type calculator will return a numeric value. Assuming that the result is '7', the natural language decorator will send the user \textit{“You did it 7 times”}, which the value '7' is wrapped in a human-readable text. 

The process of text decoration is adding the result into a predefined text prompt. In particular, the implementation of ''Answer Translator'' is  a set of rules that decides about choosing a right prompt from the list of available prompts. The rules for text prompts retrieval is based on the query keywords (described in \cite{rawassizadeh2017natural}), and the answer number is substituted. For example, if the question has "run" as sensor term and "last week" as temporal notation term, the text prompt for answer substitute a number as X in the following text prompt, "you [verb for sensor] [X] [temporal notation]".

\begin{table}[tbh]
\begin{tabular}{|l|l|}
\hline
\multicolumn{1}{|c|}{Query Type} & \multicolumn{1}{c|}{Description}                                                                                                                                    \\ \hline
keyword existence & Check the membership of a keyword in a specific period                                                                                      \\ \hline
keyword count     & Count the frequency of a keyword in a specific period                                                                                       \\ \hline
period max/min    & \begin{tabular}[c]{@{}l@{}}Calculate the maximum/minimum value in a specific period, \\ such as maximum daily steps in a month\end{tabular} \\ \hline
period sum        & \begin{tabular}[c]{@{}l@{}}Calculate the sum value in a specific period, \\ such as the total steps in a week\end{tabular}                  \\ \hline
period average    & \begin{tabular}[c]{@{}l@{}}Calculate the average value in a specific period, \\ such as the average heart rate in a one day\end{tabular}    \\ \hline
\end{tabular}
\caption{ Descriptions of five different query types that have been extracted from literature review \cite{rawassizadeh2017natural,li2014nalir}}
\label{tab:2}

\end{table}

\section{Experimental Evaluation}

In this section, we first evaluate the compression efficiency of the listed lossless compression algorithms. Next, we report the impact of using or not using the Bloom filter or Huffman coding in our approach. We then provide a comparison between state-of-the-art search methods (SOTA) and  \emph{ODSearch}. To perform this comparison, we measured query response time, memory usage, energy consumption, and scalability on smartphones and smartwatches. 

All search experiments were implemented on a mobile phone and a smartwatch. We built a mHealth conversational interface, based on UbiqLog architecture \cite{rawassizadeh2013ubiqlog} for both mobile phones and smartwatches to query a user's PHRs, and then obtain performance metrics (response time, memory utilization, and energy usage) for each experiment. 

\subsection{Experiments Setup}

The mobile phone device we used for experiments is Oukitel C12, equipped with 2 GB RAM, 16 GB Storage, 3300 mAh battery, and Mediatek MT6580M processor. The smartwatch device is TicWatch S2, which has only 512 MB RAM, 4 GB storage, 415 mAh battery, and Qualcomm Snapdragon Wear 4100 processor. The smartphone's operating system is Android OS version 8.1, and the smartwatch is wearOS version 2.32.

To cover different states of information retrieval in our experiments, we made five sample queries (see Table \ref{tab:3}), covering various question words, temporal terms, subjects, and aggregation words. Henceforth, we will refer to these queries using their alphabetical acronym (symbols in Table \ref{tab:3}). 

\begin{table}[tbh]
\begin{tabular}{|c|c|c|c|c|c|c|}
\hline
Symbol & Query                                                                                                    & \begin{tabular}[c]{@{}c@{}}Question\\ Word\end{tabular} & \begin{tabular}[c]{@{}c@{}}Temporal \\ Terms\end{tabular} & Subject                                                  & \begin{tabular}[c]{@{}c@{}}Aggregation \\ Word\end{tabular} & \begin{tabular}[c]{@{}c@{}}Query \\ Type\end{tabular}        \\ \hline
a      & \begin{tabular}[c]{@{}c@{}}How many days \\ did I run\\ this month ?\end{tabular}                        & \begin{tabular}[c]{@{}c@{}}how\\ many\end{tabular}      & \begin{tabular}[c]{@{}c@{}}this \\ month\end{tabular}     & \begin{tabular}[c]{@{}c@{}}activity \\ type\end{tabular} & none                                                        & \begin{tabular}[c]{@{}c@{}}keyword \\ count\end{tabular}     \\ \hline
b      & \begin{tabular}[c]{@{}c@{}}How many steps\\ did I walk\\ this month ?\end{tabular}                       & \begin{tabular}[c]{@{}c@{}}how\\ many\end{tabular}      & \begin{tabular}[c]{@{}c@{}}this \\ month\end{tabular}     & \begin{tabular}[c]{@{}c@{}}step \\ number\end{tabular}   & total                                                       & \begin{tabular}[c]{@{}c@{}}period \\ sum\end{tabular}        \\ \hline
c      & \begin{tabular}[c]{@{}c@{}}Did I run\\ last week?\end{tabular}                                           & \begin{tabular}[c]{@{}c@{}}did \\ I\end{tabular}        & \begin{tabular}[c]{@{}c@{}}last \\ week\end{tabular}      & activity                                                 & none                                                        & \begin{tabular}[c]{@{}c@{}}keyword \\ existence\end{tabular} \\ \hline
d      & \begin{tabular}[c]{@{}c@{}}What was my \\ average heart rate \\ today ?\end{tabular}                     & what                                                    & today                                                     & \begin{tabular}[c]{@{}c@{}}heart \\ rate\end{tabular}    & average                                                     & \begin{tabular}[c]{@{}c@{}}period \\ average\end{tabular}    \\ \hline
e      & \begin{tabular}[c]{@{}c@{}}How many steps\\ did I walk the \\ most in a day\\ this month ?\end{tabular} & \begin{tabular}[c]{@{}c@{}}how \\ many\end{tabular}     & \begin{tabular}[c]{@{}c@{}}this \\ month\end{tabular}     & \begin{tabular}[c]{@{}c@{}}step \\ number\end{tabular}   & most                                                        & \begin{tabular}[c]{@{}c@{}}period \\ max\end{tabular}        \\ \hline
\end{tabular}
\caption{Five sample queries used for experiments.}
\label{tab:3}
\end{table}

To reduce the stochasticity of experiments, each of our experiment was repeated at least five times, and the average number is reported as a result. Then, we conduct KS-Test on the result of experiments, if $p-value \geq 0.05$, we increased the number of experiments to achieve $p-value < 0.05$. For one experiment. H2 database we repeat the experiment 10 times to achieve $p-value < 0.05$. The reason to choose KS-test is that our data does not follow a normal distribution and thus we choose a non-parametric significance test \cite{rawassizadehML}.

\subsection{Compression Algorithms Comparison}
We compared five well-known lossless compression algorithms \cite{shanmugasundaram2011comparative} to encode and decode a set of sensor data. Fitness trackers deal with health information, and since the accuracy of that information cannot be compromised for performance, we avoid experimenting with lossy compression algorithms such as floating point compression.
To make our experiment dataset as stochastic as the real-world data, we synthesized random sensor data. The synthesized data was from the existing real-world data, but its sensor values were permuted. The size of the synthetic dataset was 188,889 Bytes which was made up of 30,000 records, including 10,000 entries for activity type, 10,000 entries for step number, and 10,000 entries for heart rate.

Table \ref{tab:4} reports the compression ratio, encoding, and decoding time of the five compression algorithms. The difference among compression algorithms was a trade-off between compression ratio and execution time, which means a computationally complex algorithm, such as LZW, achieved higher compression but required more resources to perform encoding or decoding.

\begin{table}[tbh]
\begin{tabular}{|c|c|c|c|c|}
\hline
\textbf{Algorithm}   & \textbf{\begin{tabular}[c]{@{}c@{}}Compression \\ Ratio (\%)\end{tabular}} & \textbf{\begin{tabular}[c]{@{}c@{}}Average \\ Encoding \\ Time (s)\end{tabular}} & \textbf{\begin{tabular}[c]{@{}c@{}}Average \\ Decoding \\ Time (s)\end{tabular}} & \textbf{\begin{tabular}[c]{@{}c@{}}Total Time of \\ Encoding and \\ Decoding (s)\end{tabular}} \\ \hline
LZW Encoding         & \textbf{397.34}  & 1.3074                 & \textbf{0.0236}   & 1.3310   \\ \hline
Arithmetic Encoding  & 187.55  & 0.0662   & 0.0714   & 0.1376   \\ \hline
Shanon-Fano Encoding & 180.57  & 0.4704 & 0.4918  & 0.9622 \\ \hline
Run Length Encoding  & 24.24  & \textbf{0.0548} & 0.0538   & 0.1086  \\ \hline
Huffman Encoding     & 187.17 & 0.0674  & 0.0338  & \textbf{0.1012}   \\ \hline
\end{tabular}
\caption{The results of applying five compression algorithms on the synthetic dataset.}
\label{tab:4}
\end{table}

LZW encoding had the highest computational complexity among the five algorithms, and achieved the highest compression ratio. Due to its complexity it required the longest time for encoding, 18 times more than Huffman encoding's time. 

Based on the results presented in Table \ref{tab:4}, we selected Huffman encoding because (i) Huffman encoding presented the best time efficiency based on the sum of encoding and decoding time, and were very close to the best result; (ii) although LZW had the best compression ratio, it is computationally too expensive for encoding, which is a significant issue on resource-constrained devices such as smartwatches; and (iii) compared with other compression methods in the experiment, Huffman encoding outperformed all in at least one aspects of compression ratio, encoding time, and decoding time.

\subsection{Bloom Filter and Huffman Coding Impact}

The backbone of our search framework is Bloom filters and compression (Huffman Coding). To demonstrate their effectiveness, we transformed \emph{ODSearch} into four different search methods and report their performances to estimate the independent impact of Bloom Filter and Huffman Coding. These four methods were: (i) Bloom filter only method, (ii) Huffman coding only method, (iii) neither Bloom Filter nor Huffman Coding (Brute force), and (iv) Both Bloom Filter and Huffman Coding (\emph{ODSearch}). 

Since \emph{ODSearch} is a combination of Bloom Filter and Huffman Coding, we can measure the impact of each through control variates or the "leave one out" method. In other words, if we compare \emph{ODSearch} with Huffman coding, we can analyze how the Bloom filter contributes to our method; and if we compare \emph{ODSearch} with the Bloom filter, we can understand how much time is required for compression. We present the brute force method as a baseline. Table \ref{tab:5} (a) and (b) reports the execution time results on a mobile phone and a smartwatch based on the experimental sample data of one user (total 10,265 records). Bloom filters' results for query \textit{b}, \textit{d}, and \textit{e} are marked as "N/A", because these queries are not membership queries. 

Note that the Bloom Filter can only answer queries \textit{a} and \textit{c}, which query for membership. Taking the average of both smartphone and smartwatch experiments on query \textit{a} and query \textit{c}, \emph{ODSearch} takes on average 0.0007 seconds more than Bloom Filter alone, but takes 72\% less time than Huffman Coding as expected since Bloom Filter is very efficient for membership query. \emph{ODSearch} only takes on average 7\% more time than Huffman coding for queries \textit{b}, \textit{d}, and \textit{e}, indicating that the extra computational burden of the Bloom filter is very insignificant. Therefore, we combine them to provide complete functionality for queries and improve their comprehensive efficiency. Clearly, \emph{ODSearch} is significantly faster than the Brute force method.

\begin{table}[b]
    \begin{subtable}{\textwidth}
        \centering
\begin{tabular}{|c|c|c|c|c|}
\hline
Query & \begin{tabular}[c]{@{}c@{}}Bloom Filter and \\ Compression (\emph{ODSearch})\end{tabular}                                 & Compression & \begin{tabular}[c]{@{}c@{}}Bloom \\ Filter\end{tabular} & \begin{tabular}[c]{@{}c@{}}No Bloom Filter \\ and \\ No Compression \\ (Brute Force)\end{tabular} \\ \hline
a     & 0.0226   & 0.0858      & 0.0224       & 1.5390                                                                                            \\ \hline
b     & 0.0348 & 0.0274      & N/A          & 0.4938                                                                                            \\ \hline
c     & 0.0100  & 0.0158      & 0.0092       & 0.2244                                                                                            \\ \hline
d     & 0.0040  & 0.0038      & N/A          & 0.0136                                                                                            \\ \hline
e     & 0.0288  & 0.0282      & N/A          & 0.5978                                                                                            \\ \hline
\end{tabular}
       \caption{Mobile phone query response time in seconds.}
    \end{subtable}
    \hfill
    \begin{subtable}{\textwidth}
        \centering
\begin{tabular}{|c|c|c|c|c|}
\hline
Query & \begin{tabular}[c]{@{}c@{}}Bloom Filter and \\ Compression (\emph{ODSearch})\end{tabular}                                & Compression & \begin{tabular}[c]{@{}c@{}}Bloom \\ Filter\end{tabular} & \begin{tabular}[c]{@{}c@{}}No Bloom Filter \\ and \\ No Compression \\ (Brute Force)\end{tabular} \\ \hline
a     & 0.0240  & 0.1204      & 0.0230 & 2.3728         \\ \hline
b     & 0.0550  & 0.0527      & N/A  & 0.9409   \\ \hline
c     & 0.0128  & 0.0268      & 0.0120    & 0.3582 \\ \hline
d     & 0.0108 & 0.0100      & N/A & 0.0816 \\ \hline
e     & 0.0593 & 0.0588      & N/A & 0.8254 \\ \hline
\end{tabular}
        \caption{Smartwatch query response time in seconds.}
     \end{subtable}
     \caption{Execution time (in seconds) while using Bloom Filter and Compression versus not using them.}
     \label{tab:5}
\end{table}

\subsection{Comparison with State-of-The-Art}
This section reports execution time, memory utilization, and energy usage of our approach compared to four State-of-the-Art (SOTA) methods. Three SOTA databases that are used on mobile or wearable devices were selected, including SQLite, Realm and H2. Lucene was also selected even though it is not a database itself, but because it is a popular indexing mechanism that is implemented on mobile devices \cite{bialecki2012apache,chen2012mobile,coelho2021semantic}.\\

SQLite and H2 are relational databases and it is recommended to store semantic data (in our case sensor data) in separate tables. Therefore, we use two tables one for heart rate and one for physical activities. RealM is a document database (MongoDB for smartphones) and can handle data from heterogeneous resources. It is advised to keep the data in a single data dump for document databases \cite{kleppmann}. We combine all the data for RealM into a single data dump. We use the default settings of the applications, i.e. two separate folders, one for heart rate files and one for activity files, for Lucene, ODSearch and Brute Force.

\subsection{Execution Time}
To enable \emph{ODSearch} implementation on real-world mHealth applications, an essential factor that directly affects the user experience is the execution time (a.k.a. response time) \cite{tolia2006quantifying, rawassizadeh2015lesson}. The execution time starts when the user sends a query and ends when the user receives the answer. As described above, we used SQLite, Realm, H2, and Lucene. Additionally, we added a brute force search as a baseline.

\begin{figure*}
\begin{center}
\includegraphics[scale=0.49]{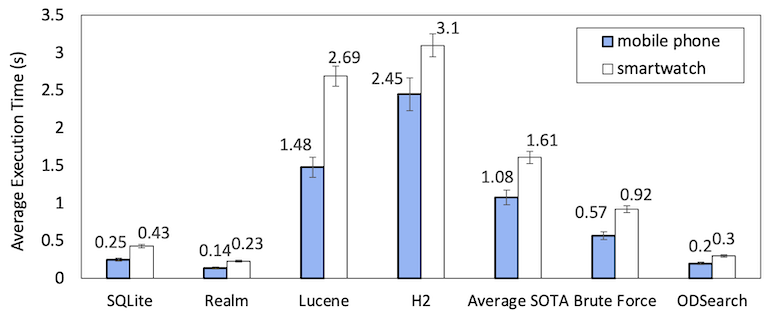}
\caption{Average execution time (in seconds) of querying the experimental sample data of one user with five queries}
\label{fig:5}
\end{center}
\end{figure*}

The mobile phone and smartwatch execution time results are shown separately in Table \ref{tab:6} (a) and (b). In addition, Figure \ref{fig:5} presents the average execution time of the five queries from Table \ref{tab:6}. Results in Figure \ref{fig:5} shows that our approach significantly outperforms all other search mechanisms on both devices. For both smartphone and smartwatch, on average, the execution time is \textit{55 times faster} than SOTA methods and \textit{6 times faster} than the fastest SOTA method, i.e., Realm. 

\begin{table} 
    \begin{subtable}{\textwidth}
        \centering
\begin{tabular}{|c|c|c|c|c|c|c|}
\hline
Query & SQLite & Realm & Lucene & H2 & Brute Force & \textbf{ODSearch} \\ \hline
a              & 0.6071          & 0.2214         & 3.5082          & 4.3263              & 1.5390      & \textbf{0.0226} \\ \hline
b              & 0.2263          & 0.1910         & 1.5011          & 3.3001              & 0.4938      & \textbf{0.0348} \\ \hline
c              & 0.1713          & 0.0318         & 0.7646          & 1.1672              & 0.2244      & \textbf{0.0100} \\ \hline
d              & 0.0181          & 0.0086         & 0.0524          & 0.0650              & 0.0136      & \textbf{0.0040} \\ \hline
e              & 0.2412          & 0.2556         & 1.5832          & 3.4228              & 0.5978      & \textbf{0.0288} \\ \hline
\end{tabular}
       \caption{Mobile Phone}
    \end{subtable}
    \hfill
    \begin{subtable}{\textwidth}
        \centering
\begin{tabular}{|c|c|c|c|c|c|c|}
\hline
Query & SQLite & Realm & Lucene & H2 & Brute Force & \textbf{ODSearch} \\ \hline
a              & 0.783          & 0.3923         & 6.5570          & 5.66187              & 2.3728      & \textbf{0.0240} \\ \hline
b              & 0.5820          & 0.2874         & 2.6058          & 4.1271              & 0.9409      & \textbf{0.0550} \\ \hline
c              & 0.2521          & 0.1532         & 1.4270          & 1.2914              & 0.3582      & \textbf{0.0128} \\ \hline
d              & 0.0366          & 0.0296         & 0.0650          & 0.0361              & 0.0816      & \textbf{0.0108} \\ \hline
e              & 0.4959          & 0.3016         & 2.7794          & 4.7690              & 0.8254      & \textbf{0.0593} \\ \hline
\end{tabular}
        \caption{Smartwatch}
     \end{subtable}
     \caption{Execution time (in seconds) of querying the experimental sample dataset.}
     \label{tab:6}
\end{table}

Although it is not common to report, we also report the execution time of the preprocessing for each search method. Table \ref{tab:7} (a) and (b) show the average preprocessing time for each search method. Brute Force does not have a preprocessing step, therefore its preprocessing result is almost zero. We suspect the overhead of 0.2 seconds originates from the Android and WearOS operating systems internal processes. Except for the Brute Force method, preprocessing of \emph{ODSearch} ranked second among mobile phone experiments and third among smartwatch experiments. However, the difference is negligible because preprocessing will rarely be done \cite{rawassizadeh2019indexing}, e.g. once a day or when the device is charging and there is no interaction with the device.

\begin{table}[H]
    \begin{subtable}{\textwidth}
        \centering
\begin{tabular}{|l|c|c|c|c|c|c|}
\hline
   & SQLite & Realm & Lucene & H2 & Brute Force & ODSearch \\ \hline
\multicolumn{1}{|c|}{Time} & 30.0   & \textbf{5.5}   & 12.8   & 15.2 & \textbf{0.2}  & 7.3      \\ \hline
\end{tabular}
       \caption{Mobile Phone}
    \end{subtable}
    \hfill
    \begin{subtable}{\textwidth}
        \centering
\begin{tabular}{|l|c|c|c|c|c|c|}
\hline
& SQLite & Realm & Lucene & H2 & Brute Force & ODSearch \\ \hline
\multicolumn{1}{|c|}{Time} & 249.9  & \textbf{18.8}  & 44.5  & 26.0 & \textbf{0.2}  & 42.0     \\ \hline
\end{tabular}
        \caption{Smartwatch}
     \end{subtable}
     \caption{ Average preprocessing time (in seconds) of the experimental sample data of one user}
     \label{tab:7}
\end{table}

\subsection{Memory Utilization}

In comparison to conventional computers, another limited resource in mobile or wearable devices is memory \cite{bhattacharya2016sparsification}. We use the memory profiler\footnote{https://developer.android.com/studio/profile/memory-profiler} in Android Studio to collect the memory usage of the device in real-time. Memory profiler identified that the preprocessing of all search tools required the most amount of memory.

\begin{figure}[htb]
\begin{center}
\includegraphics[scale=0.41]{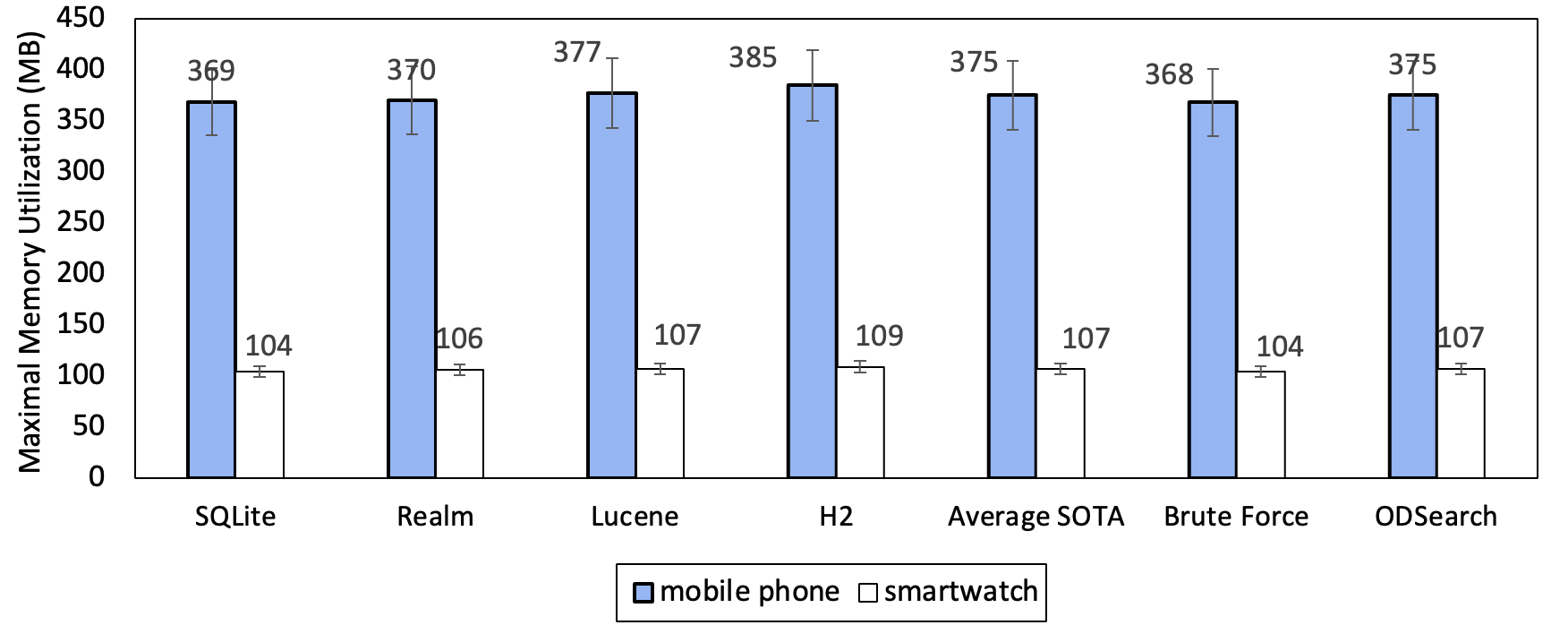}
\caption{Memory utilization (in MegaBytes) of preprocessing the sample dataset}
\label{fig:6}
\end{center}
\end{figure}

Figure \ref{fig:6} presents the memory utilization results on mobile phones and smartwatches separately for the preprocessing phase. In both mobile phone and smartwatch experiments, \emph{ODSearch} outperformed all other SOTA methods in terms of memory usage. Compared with the average performance of SOTA methods, \emph{ODSearch} requires \textit{1.9\% less memory} on the mobile phone and \textit{2.7\% less memory} on the smartwatch. Furthermore, \emph{ODSearch} takes \textit{0.2\% less memory} on the mobile phone and \textit{0.4\% less memory} on the smartwatch than SQLite, which is the most memory-efficient SOTA method.

\subsection{Energy Usage}

Mobile and wearable devices are known for their limited battery capacity \cite{rawassizadeh2015energy, homayounfar2020understanding} and energy utilization directly affects the usability of the device \cite{visuri2017quantifying}. We report the energy usage for query execution to identify the energy impact of \emph{ODSearch}. Energy usage monitoring started when a query input was entered into the GUI and ended when the user received the answer. To monitor the battery usage in millijoules ($mJ$) or joules ($J$), we analyzed the battery changes at its current ampere ($A$) and voltage ($V$). Since we know the interval of a query or reprocessing (in second, $S$), we can compute the energy consumption based on the following equation: $1\;J( 1000\;mJ) = 1\;V \times 1\;A \times 1\;S$ 

\begin{figure*}[h]
\begin{center}
\includegraphics[scale=0.37]{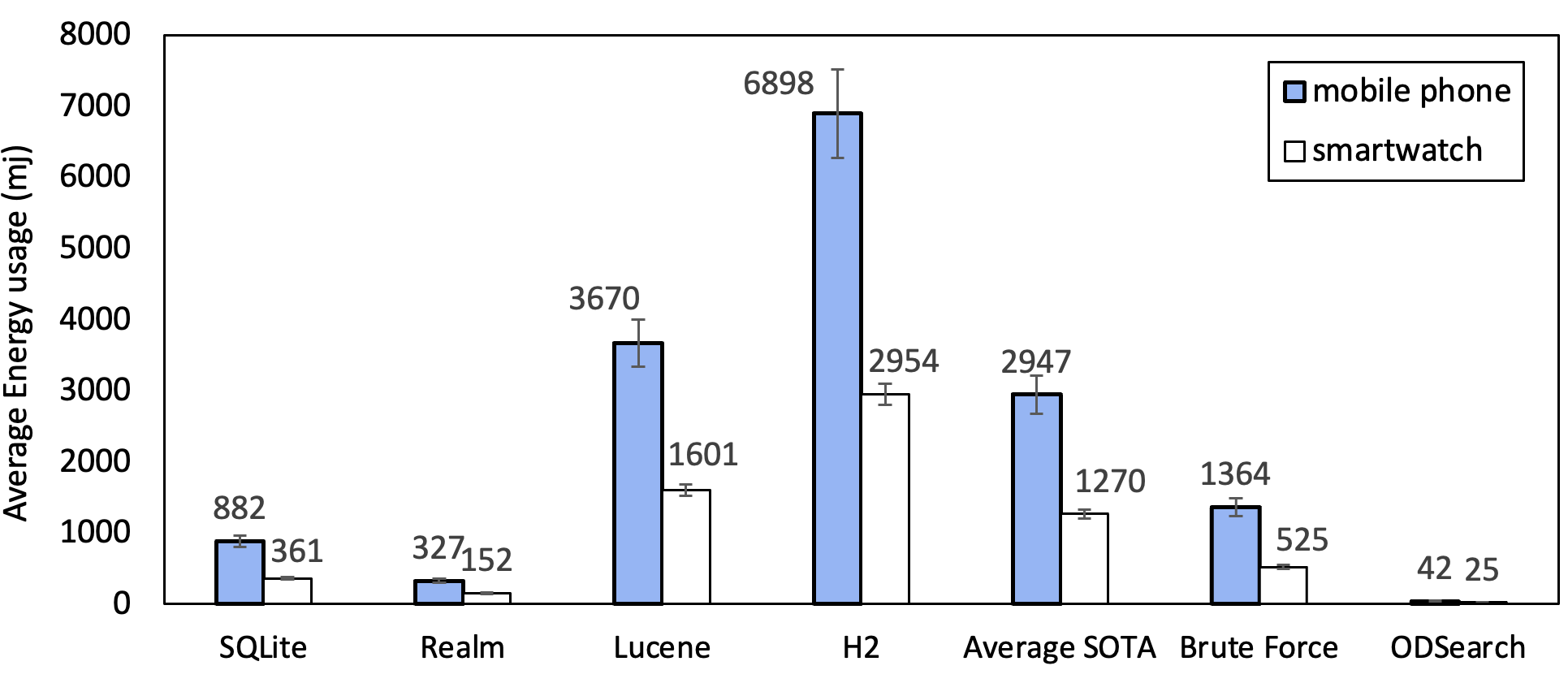}
\caption{Average energy usage (in mJ) with five queries}
\label{fig:7}
\end{center}
\end{figure*}

Table \ref{tab:8} shows the average energy usage in millijoules for the five repeated queries, on both devices. Moreover, Figure \ref{fig:7} presents the average results of the five sample queries from Table \ref{tab:8}. It is evident that \emph{ODSearch} outperforms all other methods, in terms of energy use. In particular on mobile phones, the SOTA methods \textit{consume 20 times more energy} on average than \emph{ODSearch}, and this difference increases \textit{to 32 times} on the smartwatch. Compared with \emph{ODSearch}, the most energy-efficient method was Realm, however, \emph{ODSearch} consumes \textit{7 times less energy} than Realm on the mobile phone and \textit{5 times less energy} on the smartwatch.

\begin{table}[H]
    \begin{subtable}{\textwidth}
        \centering
        \begin{tabular}{|c|r|r|r|r|r|r|}
\hline
Query & SQLite  & Realm & Lucene  & H2 & Brute Force & \textbf{ODSearch}       \\ \hline
a     & 2,116.2 & 539.8 & 8,797.8 & 13,471.0   & 4,007.8     & \textbf{226.0} \\ \hline
b     & 833.8   & 428.2 & 4,174.8 & 9,024.6    & 1,083.8     & \textbf{82.6}  \\ \hline
c     & 440.0   & 73.8  & 1,741.4 & 3,075.0    & 537.8       & \textbf{26.2}  \\ \hline
d     & 33.4    & 20.0  & 119.6   & 12.2       & 27.2        & \textbf{9.0}   \\ \hline
e     & 1,028.2    & 586.0  & 3,515.8   & 8,908.0      & 1,164.8       & \textbf{50.6}   \\ \hline
\end{tabular}
       \caption{Mobile Phone}
    \end{subtable}
    \hfill
    \begin{subtable}{\textwidth}
        \centering
\begin{tabular}{|c|r|r|r|r|r|r|}
\hline
Query & SQLite & Realm & Lucene  & H2 & Brute Force & \textbf{ODSearch}      \\ \hline
a     & 827.0  & 274.0 & 3,646.3 & 4,896.6    & 1,142.2     & \textbf{88.0} \\ \hline
b     & 417.9  & 193.3 & 1,572.6 & 4,326.8    & 651.6       & \textbf{47.5} \\ \hline
c     & 137.8  & 93.4  & 1,134.0 & 1,117.8    & 202.0       & \textbf{13.3} \\ \hline
d     & 31.6   & 21.2  & 41.0    & 27.4       & 45.0        & \textbf{7.4}  \\ \hline
e     & 425.0   & 208.6  & 1,609.0    & 4,403.8       & 582.8        & \textbf{36.4}  \\ \hline
\end{tabular}
        \caption{Smartwatch}
     \end{subtable}
     \caption{Energy usage (in mJ) of querying experimental sample data of one user}
     \label{tab:8}
\end{table}

\subsection{Scalability}
To measure scalability, we examined how the performance of the different search methods changes as the data size increased. We used synthetic datasets with increasing sizes for this experiment. As discussed, the synthetic datasets were constructed from the real-world data of users, but the sizes of synthetic datasets varied, i.e., 30kB, 3MB, 6MB, 12MB, 18MB, 24MB, 30MB, 36MB, 42MB, 48MB. 

\subsubsection{Scalability on Mobile Phone}
First, we measured the memory utilization of preprocessing for each data size. If the data size was too large, the preprocessing would be terminated and is shown as an "ERROR" representing Java “Out Of Memory Error” java.lang.OutOfMemoryError in our report. Table \ref{tab:9} presents the memory utilization for each method. The Brute Force method does not have a preprocessing step, hence it is excluded. Our experiments reveal that \emph{ODSearch} can preprocess a maximum 42 MB of data at once on the experimental smartphone. We investigated the reason for this and recognized that the size of the Bloom filter causes the Java heap memory to get full and raised the error. If a particular application intends to use ODSearch for data larger than the size of 42MB, the simplest way to fix this issue is to write the Bloom filter's contents to disk. Though reading from the disk is slower than reading from memory, there will still be extra overhead associated with it.
There are other common approaches, such as increasing the Java heap size for ODsearch process. To keep the experiment fair among all methods we do not experiment with them.

Although both SQLite and Lucene could preprocess 48MB dataset, for all datasets less than 42MB, \emph{ODSearch} required the least amount of memory, as shown in Table \ref{tab:9}.

\begin{table}[tbh]
\centering
\begin{tabular}{|c|c|c|c|c|c|}
\hline
\multicolumn{1}{|c|}{Data   Size} & \multicolumn{1}{c|}{SQLite} & \multicolumn{1}{c|}{Realm} & \multicolumn{1}{c|}{Lucene} & \multicolumn{1}{c|}{H2} & \multicolumn{1}{c|}{ODSearch} \\ \hline
30Kb & 374.1 & 375.0 & 373.1 & 377.3 & \textbf{373.0} \\ \hline
3Mb & 375.4 & 376.7 & 378.1 & 395.4 & \textbf{375.1} \\ \hline
30Mb & 448.5 & 410.7 & 443.3 & 418.0 & \textbf{382.9} \\ \hline
36Mb & 455.6 & 413.0 & 455.6 & 433.7 & \textbf{390.3} \\ \hline
42Mb & 483.4 & Error & 468.9 & 449.1 & \textbf{397.5} \\ \hline
48Mb & 495.5 & Error & \textbf{481.3} & Error & Error \\ \hline
\end{tabular}
\caption{Maximum memory utilization (in MB) of preprocessing synthetic dataset on the mobile phone used for our experiments.}
\label{tab:9}
\end{table}
Our second objective was to determine the marginal increase in memory usage as data size increased. In other words, how much extra memory would be required when we have to preprocess 1 MB \footnote{Based on our experiment 1MB of data is an average data for one week and existing fitness trackers are usually presenting data for one week.} more data? To answer this question we calculated the average memory usage for data sizes between 30 KB and the maximum loadable datasets for each search method. To preprocess an extra 1 MB of data, \emph{ODSearch} only required 0.58 MB memory, while SQLite required 2.53 MB memory, Realm required 1.06 MB memory, Lucene required 2.26 MB memory, and H2 required 1.71 MB memory. Therefore, \emph{ODSearch} outperforms all other methods in terms of marginal memory utilization.

\begin{figure*}[h]
\begin{center}
\includegraphics[scale=0.4]{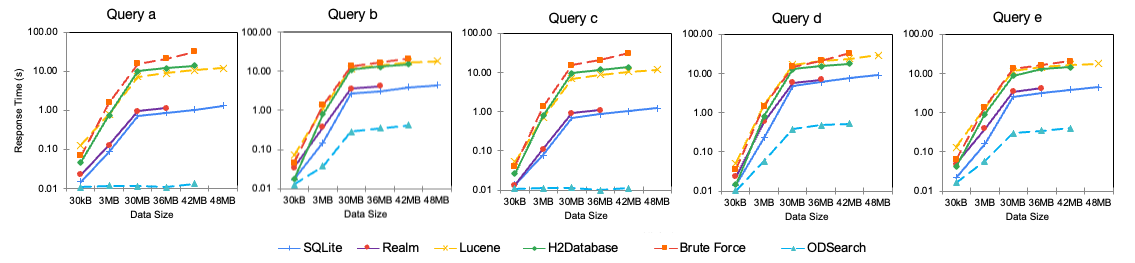}
\caption{Response time of querying synthetic data on mobile phone. For clarity the Y axis is presented in logarithmic scale. Note that the intervals between dataset sizes on X axis are not equal.}
\label{fig:8}
\end{center}
\end{figure*}

We also explored the changes in response time as a function of the size of the dataset. Since the difference among the query response times is too large to visualize, we report the response times for the five sample queries with logarithmic scale (see Figure \ref{fig:8}). In all of these experiments, \emph{ODSearch} outperforms all other methods. Another advantage of \emph{ODSearch} is its constant efficiency as illustrated by queries \textit{a} and \textit{c} (see Figure \ref{fig:8}). This is because both queries \textit{a} and \textit{c} are membership queries and \emph{ODSearch} uses Bloom filter to query membership with $O(1)$ complexity for our synthetic datasets.

\begin{figure*}[h]
\begin{center}
\includegraphics[scale=0.53]{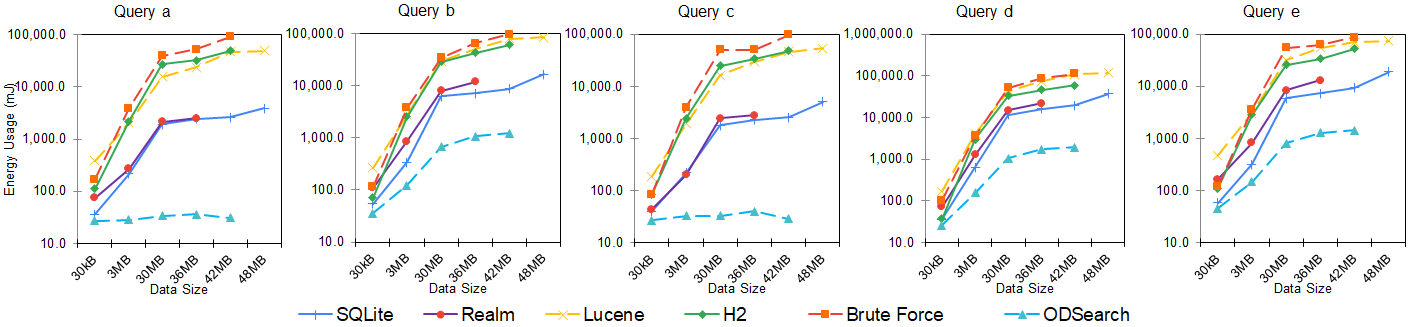}
\caption{Energy usage of querying synthetic data on mobile phone. For clarity the Y axis is presented in logarithmic scale. Note that the intervals between dataset sizes on X axis are not equal.}
\label{fig:9}
\end{center}
\end{figure*}

Moreover, we report the energy usage for querying different sizes of data. Since the energy usage also differs for different dataset sizes, we use a logarithmic scale to present five charts for the five sample queries in Figure \ref{fig:9}. \emph{ODSearch} consumed the least amount of energy and also outperformed all other methods in terms of energy consumption.

\subsubsection{Scalability on Smartwatch}

Table \ref{tab:10} reports the maximum memory used for each search tool on the smartwatch. SQLite, Realm, and H2 cannot preprocess 30 MB datasets. Conversely, although both \emph{ODSearch} and Lucene have the preprocessing capacity of at least 30 MB, \emph{ODSearch} required \textit{19\% less memory} than Lucene, indicating that \emph{ODSearch} outperforms all other methods on memory utilization on smartwatches.

\begin{table}[H]
\centering
\begin{tabular}{|c|r|r|r|r|r|}
\hline
\multicolumn{1}{|c|}{Data Size} & \multicolumn{1}{c|}{SQLite} & \multicolumn{1}{c|}{Realm} & \multicolumn{1}{c|}{Lucene} & \multicolumn{1}{c|}{H2} & \multicolumn{1}{c|}{ODSearch} \\ \hline
30kB  & 103.8  & 109.0 & 103.7 & 110.0 & \textbf{112.3} \\ \hline
3MB & 116.9 & 115.4 & 110.6 & 115.0 & \textbf{116.0} \\ \hline
6MB & 121.1 & 120.4 & 120.6 & 137.9 & \textbf{120.1} \\ \hline
12MB & 133.3 & 125.7 & 131.3 & 148.1 & \textbf{123.2} \\ \hline
18MB & 145.4 & 130.8 & 142.9 & 158.3 & \textbf{126.2} \\ \hline
24MB & 157.2 & 136.6 & 153.6 & 168.8 &\textbf{128.7} \\ \hline
30MB & \multicolumn{1}{r|}{Error} & \multicolumn{1}{r|}{Error} & 163.4 & \multicolumn{1}{r|}{Error} & \textbf{132.4} \\ \hline
36MB & \multicolumn{1}{r|}{Error}  & \multicolumn{1}{r|}{Error} & \multicolumn{1}{r|}{Error}  & \multicolumn{1}{r|}{Error}  & \multicolumn{1}{r|}{Error}    \\ \hline
\end{tabular}
\caption{Maximum memory utilization (in MB) of preprocessing synthetic dataset on smartwatch. “Error” here refers to Java “Out of Memory Error”.}
\label{tab:10}
\end{table}

We also report marginal memory utilization on a smartwatch. These results are calculated between their maximal loadable synthetic dataset (30 MB or 24 MB) and 30 KB datasets. To preprocess an extra 1 MB of data, \emph{ODSearch} required the least amount of memory, 0.67 MB, while SQLite required 2.23 MB, Realm required 1.15 MB, Lucene required 1.99 MB, and H2 required 2.45 MB. Therefore, \emph{ODSearch} outperforms all other methods in terms of marginal memory utilization. All methods failed to search 36MB of data, We can use the methods we previously discussed for the smartphone as a solution.
\begin{figure*}[h]
\begin{center}
\includegraphics[scale=0.53]{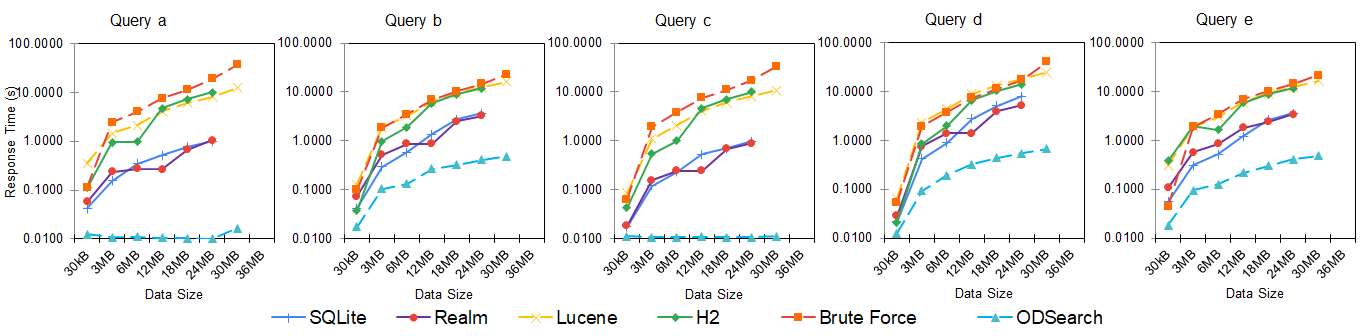}
\caption{Response time of querying synthetic data on smartwatch. For clarity the Y axis is presented in logarithmic scale. Note that the intervals between data sizes on X axis are not equal.}
\label{fig:10}
\end{center}
\end{figure*}

\begin{figure*}[h]
\begin{center}
\includegraphics[scale=0.53]{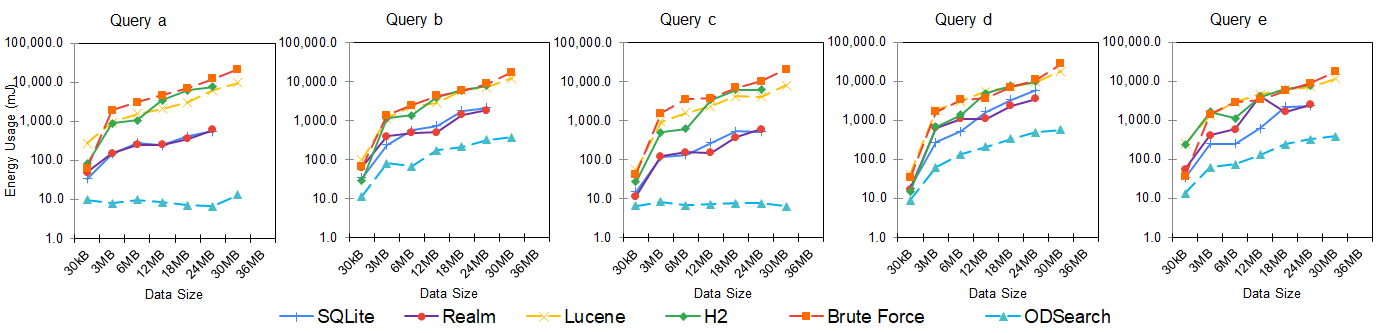}
\caption{ Energy usage of querying synthetic data on smartwatch. For clarity the Y axis is presented in logarithmic scale. Note that the intervals between data sizes on X axis are not equal.}
\label{fig:11}
\end{center}
\end{figure*}
Figure \ref{fig:10} presents the query response time on a smartwatch, and Figure \ref{fig:11} presents the energy usage for the five sample queries. According to the response time and energy utilization results, \emph{ODSearch} is the most energy-efficient and the fastest method. These results also verify that, even on the smartwatch, \emph{ODSearch} can answer a query between 0.01 seconds and 0.7 second, which can be considered as a real-time response.

\section{Discussion}

Here we briefly describe the limitations of our approach and how it can contribute to improving the existing fitness tracker applications and devices.  

We analyze a longitudinal wearable and smartphone data.  Few studies have addressed the issue of using longitudinal wearable data for medical purposes \cite{rohani2018correlations, perez2019large}. However, unlike our work, they do not use large data to improve user experiences; instead, they analyze user data offline for other purposes. To the best of our knowledge, there aren't many works that use longitudinal wearable data to enhance user experience (visuri2017quantifying, visuri2021understanding). In this case, longitudinal data help us enhance the user experience. Nevertheless, the emphasis of this work is on the technical specifications of a novel on-device search component that is proposed. Our future work will consist of analyzing its effect on the user interface.   \\
Studies have shown that conversational agents have advantages over conventional graphical user interfaces in applications like mental health \cite{gaffney2019, vaidyam2019}. Especially while dealing with fitness tracker applications, conversational agents provide more motivation for behavior changes in comparison to traditional interfaces  \cite{kocielnik2018}. However, they are not adapted to industrial fitness trackers. We checked a number of well-known fitness trackers at the time we were writing this article,  Google Fit \footnote{https://www.google.com/fit}, Apple Health \footnote{https://www.apple.com/ios/health}, FitBit \cite{https://www.fitbit.com/global/us/setup}, Strava \footnote{https://www.strava.com/mobile} and Samsung Health \footnote{https://www.samsung.com/global/galaxy/apps/samsung-health}. They all use browsing and information visualization, and they do not provide conversational agents. They don't all offer conversational agents; instead, they all rely on browsing and information visualization. With the exception of Apple Health, they transfer users' data into the cloud and then delete them from the device.

An interface that gives users retrospective data is necessary for conversational agents to function on par with fitness trackers' graphical user interfaces. This need is met by ODSearch, which offers a quick and resource efficient search component along with other features like network independence.

The sparsity of data, which affects accuracy, is a common problem when working with consumer electronic wearables. The origin of sparsity is varied, including operating systems that kill background processes \footnote{https://dontkillmyapp.com [last visit 16-Jul-2022]}, inaccurate sensor reading, \cite{chong2020consumer}, users' manually removing the device from their body, \cite{rawassizadeh2015lesson}, users' removing the device from their proximity, \cite{dey2011getting}. The dataset used in this study is taken from a real-world dataset, and these problems were present in that dataset. As a result, the data gathered might appear to be less extensive than a reader might initially assume given their expectation of data availability 24/7.

Another question that the reader might have is, "Why haven't we used state-of-the-art question-answering systems such as BERT \cite{devlin2018} or RoBERTa \cite{2019roberta} that rely on big language models?" We have three justifications for not using language models. First, we have experimented with state-of-the-art question-answering methods, despite their high accuracy, they are slow, and thus not feasible to implement in a real-time application. Second, they work best with plain text rather than tabular or hierarchical data. The data collected from sensors is in hierarchical forms and is not a proper fit for language models. Third, previous research shows \cite{rawassizadeh2017natural} a limited number of questions could be asked about health and thus we are dealing with closed domain questions. Language models are useful to implement open domain questions, but currently, they are too heavy and resource intensive to get integrated into small battery powered devices such as smartwatches. 

Future work will concentrate on developing a conversational agent based on \emph{ ODSearch}, and we'll carry out a user study to assess the effects of doing so as opposed to the conventional graph-based data representation that is currently common in fitness applications.

\section{Conclusion}

In this work, we describe \textit{ODSearch}, a network-independent, fast, and resource-efficient search framework that interacts with natural language and answers queries on mobile phones and wearables in real-time or near real-time. Our method consists of four modules, of which \textit{Query Translator} and \textit{Answer Translator} modules are designed for natural language interaction, and \textit{Bloom Filter} and \textit{Huffman Coding} modules are built for efficient information retrieval. We have experimented with several lossless compression algorithms and demonstrate that Huffman coding is the best compression algorithm for \emph{ODSearch}, and it can significantly improve the execution time of the search operation.  Furthermore, we present that \emph{ODSearch} can query in real-time on mobile phone and smartwatch devices and outperform state-of-the-art search methods on large or small datasets, in execution time, energy utilization and memory use. Additionally, \emph{ODSearch} is among the most scalable search methods that can handle large data searches, while consuming the minimum amount of memory and energy. 

\bibliographystyle{ACM-Reference-Format}
\bibliography{sample-base}

\appendix

\end{document}